\documentstyle[aps,multicol,ifthen,epsf]{revtex}

\newcommand{\be}{\begin{equation}}
\newcommand{\ee}{\end{equation}}
\newcommand{\ba}{\begin{eqnarray}}
\newcommand{\ea}{\end{eqnarray}}

\begin{document}
\draft

\title{Self-organized critical random directed polymers}
\author{Per J\"ogi$^{1,2,3}$ and Didier Sornette$^{2,3}$}
\address{
   $^1$ Department of Physics,
        University of California, Los Angeles, California 90095-1567 \\
   $^2$ Institute of Geophysics and Planetary Physics \\and
        Department of Earth and Space Sciences\\
        University of California, Los Angeles, California 90095-1567 \\
   $^3$ Laboratoire de Physique de la Mati\`{e}re Condens\'{e}e\\
   		CNRS and Universit\'{e} de Nice-Sophia Antipolis, Parc Valrose,
        06108 Nice, France
}

\maketitle

\begin{abstract}
We uncover a nontrivial signature of the hierarchical structure of
quasi-degenerate random directed polymers (RDPs) at zero temperature in
$1+1$ dimensional lattices.  Using a cylindrical geometry with
circumference $8 \leq W \leq 512$, we study the differences in
configurations taken by RDPs forced to pass through points displaced
successively by one unit lattice mesh.  The transition between two
successive configurations (interpreted as an avalanche) defines an 
area $S$.  The distribution of moderatly sized avalanches is found to be 
a power-law $P(S)\,dS \sim S^{-(1+\mu)}\,dS$.  Using a hierarchical 
formulation based on the length scales $W^{2\over 3}$ (transverse excursion) 
and the distance $W^{{2\over 3}\alpha}$ between quasi-degenerate ground
states (with $0<\alpha\le 1$), we determine $\mu = {2\over 5}$, in
excellent agreement with numerical simulations by a transfer matrix method.
This power-law is valid up to a maximum size $S_{5\over 3} \sim W^{5\over 3}$.
There is another population of avalanches which, for characteristic sizes 
beyond $S_{5\over 3}$, obeys $P(S)\,dS \sim \exp(-(S/S_{5\over 3})^3)\,dS$ 
also confirmed numerically.  The first population corresponds to almost
degenerate ground states, providing a direct evidence of ``weak replica 
symmetry breaking'', while the second population is associated with different
optimal states separated by the typical fluctuation $W^{2\over 3}$ of a 
single RDP.


\end{abstract}


\begin{multicols}{2}
\narrowtext

\section{Introduction}
\label{sec:intro}

Self-organized criticality (SOC) \cite{Bak} describes out-of-equilibrium
extended systems driven infinitely slowly which respond intermittently with
avalanches or bursts of sizes distributed according to power-law distributions.
A close relationship between critical phase transitions and a class of SOC
systems \cite{Baktang,Mapping} has been pointed out.  Member systems of
this SOC class operate exactly at the critical value of an underlying critical
point.   A necessary condition for this to occur is that the {\it order
parameter} (often akin to a flux) of a dynamical critical transition be driven
infinitely slowly, thus forcing the control parameter to readjust itself
dynamically around its critical value \cite{Mapping}.

Motivated by this correspondence, we introduce a new SOC model.  It can be
described as an {\it equilibrium} depinning problem wherein a certain type of
avalanche separates local equilibrium states.  The succession of equilibrium
state transitions found in our model resembles the behaviour of abelian
sandpiles \cite{Dhar}.  In the latter, each avalanche can be shown to connect
two different microscopic metastable states.  Furthermore, its critical state
is then characterized by the complete set of these avalanche-connected
metastable states.  Whereas the set of coexisting metastable stables is
created by the threshold rules of the sandpile automata, the many coexisting
local {\it equilibrium} states appearing in our model emerge from an optimal
(i.e. minimum energy) configuration in a quenched random landscape.  This
disorder induces the coexistence of an extremely large number of almost
equivalent configurations.  The resulting closeness in energy space leads
to a large spread in configuration space.  This turns out to produce a
power-law distribution for the interconnecting avalanches.

Thus, a common property of SOC systems is that they are characterized by
a large set of almost equivalent and degenerate states.  This set can be
generated by dynamic automata rules, disorder, frustration or other mechanisms.
In addition to the introduction of a new class of SOC models, our results
provide further evidence for the hierarchical structure of sets of random
directed polymers (RDPs).

Our results bear an apparent strong similarity to those previously obtained
for pinned charged density waves \cite{middle}, driven interfaces in random
media \cite{Nofish}, and elastic manifolds on disordered substrates \cite{Hwa}.
However, the connection between the dynamic critical phenomena obtained from 
a constant driving force $F$ at the depinning threshold $F_c$ and the nearly
critical behavior obtained by a small constant velocity drive is based on an
argument relating the critical behavior as $F \to F_{c}^{+}$ and 
$F \to F_{c}^{-}$.  This predicts \cite{middle,Nofish,Hwa} a vanishing exponent
for the avalanche distribution in our 1+1 dimensional case, which seemingly
contradicts our result.  The discrepancy stems from the fact that we do not
describe the same regime; the vanishing exponent refers to the existence of
large avalanches of sizes controlled by the system size (or the correlation
length when off-criticality applies).  This corresponds to the second of two
identified avalanche regimes of our model.  In contrast, the present work
reveals the existence of a sub-dominant power-law distribution of avalanches
stemming from the hierarchy of almost equivalent degenerate states.  These
states do not, however, contribute to the large scale behavior and have thus
been overlooked in previous work.

The model is defined in the next section, while in Sec.~\ref{sec:thpr} we 
derive our theoretical predictions for the distribution of avalanche sizes.
These are compared with extensive numerical simulations in
Sec.~\ref{sec:numtst}.  Our conclusions are found in Sec.~\ref{sec:conc}.
\end{multicols}\widetext\noindent
\begin{figure}
  \centerline{\mbox{\epsfbox[0 0 509 360]{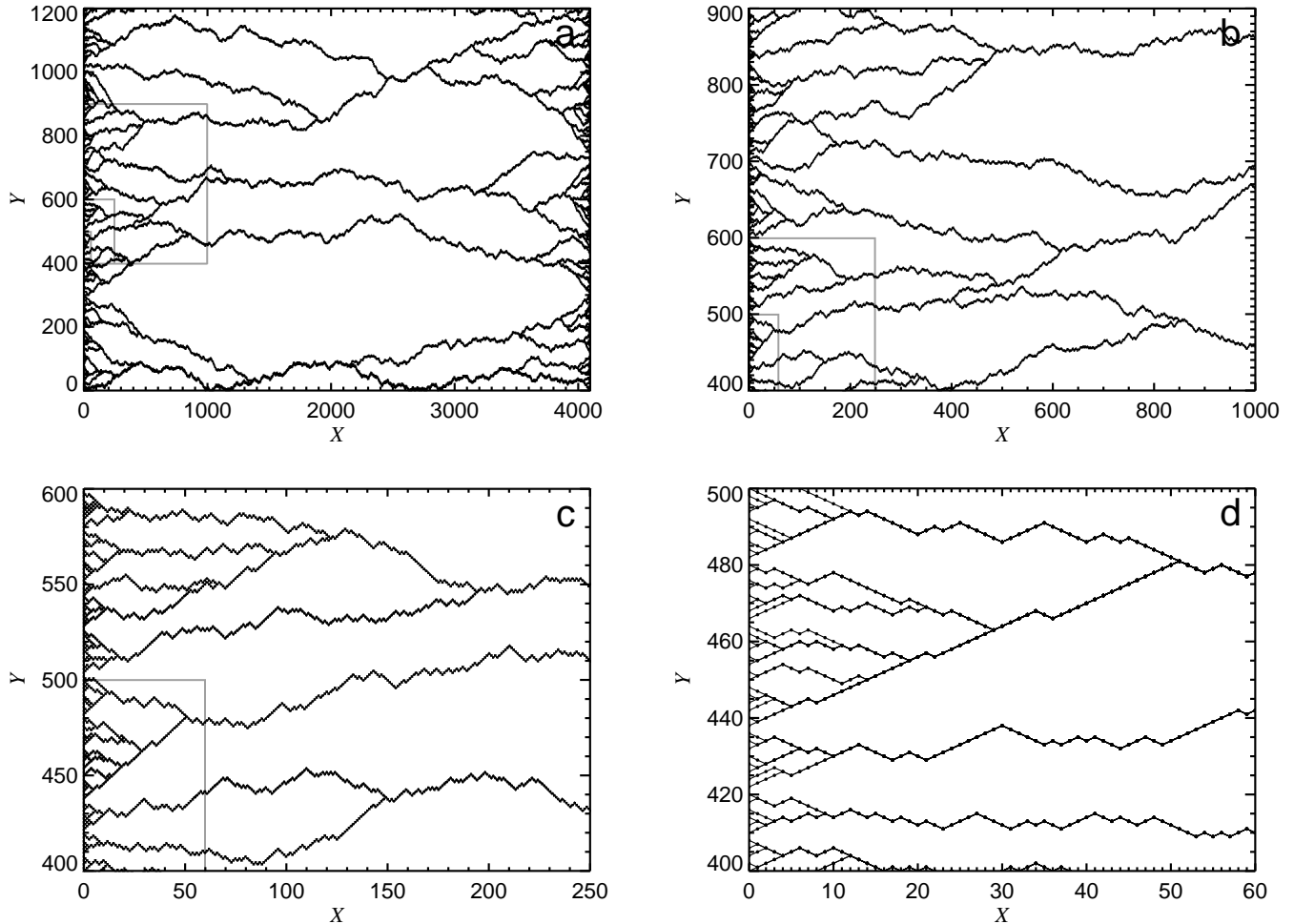}}}
  \caption{
           A typical set of optimal configurations for a RDP of length
           $W=4096$ and for $0 \leq y \leq 1200$.
           a) global system (grey framed boxes outlines regions of
           succeeding plots such that the horizontal and vertical extensions
           of these boxes follow Eqs.~(\ref{eq:lnpln}) and (\ref{eq:wnpln})
           with $\alpha\approx 0.9$);
           b) magnification of the largest box in a);
           c) magnification of the largest box in b);
           d) magnification of the box in c).
          }
  \label{fig1}
\end{figure}
\begin{multicols}{2}\narrowtext
\noindent

\section{Definition of the model}
\label{sec:defmo}

Consider a RDP on a square lattice oriented at $45^\circ$ with respect to 
the $x$ axis and such that each bond carries a random number, interpreted 
as an energy.  An arbitrary directed path (a condition of no backwards turn)
along the $x$-direction and of length $W$ (in this direction) corresponds 
to the configuration of a RDP of $W$ bonds.  In the zero temperature version 
we study here, the equilibrium polymer configuration is the particular directed
path on this lattice which (in the presence of given boundary conditions)
minimizes the sum of the $W$ bond energies along it.  This simple model, 
with its much varied behavior, has become a valuable tool in the study of
self-similar surface growths \cite{KPZ}, interface fluctuations and depinning
\cite{Huse}, the random stirred Burgers equation in fluid dynamics \cite{Kardar}
and the physics of spin glasses \cite{spinglass}.

Let us apply a field $h$ that exerts a force on one of the vertical endpoint
position $y(W)$ of the polymer.  This field adds a term $-h\,y(W)$ to the
configurational energy of the polymer given by the sum of random bond energies
along it.  It is similar to a transverse electric field acting on the charged
head of the polymer.  If the other polymer extremity is free, the minimum
energy is obtained by letting $y(W)$ go to infinity as the  external field
term $-h\,y(W)$ diverges to $-\infty$.  This energy always dominates the
configuration energy for any reasonable distribution of random bond energies.
A depinning transition thus occurs for the value $h=0^{+}$ of the control
parameter $h$.  M\'ezard \cite{Mezard} has shown that holding the other
endpoint fixed results (in the small field limit) in extremely jerky
displacement of the charged head as a function of the field strength $h$.  
The position of the charged head is stationary for large ranges of applied 
field values and then changes suddenly.  At the field values where these
transitions (or avalanches) occur, the susceptibility attains large values.
These susceptibility bursts are reportedly distributed according to a power-law
\cite{Mezard}.  This avalanche response has been attributed \cite{Mezard} to
a ``spin glass phase'', with several valleys of similar energy.  It is
important to realize that this avalanche behavior is not SOC as the driving
is nonstationary; nothing occurs when the field stays constant and increasing
the field will lead ultimately to the situation where the RDP is blocked in
a fully extended configuration along the first quadrant bisectrix.  This
regime is similar to a mode of operation with a slow sweeping of a control
parameter \cite{Sweeping}.
\begin{figure}
  \centerline{\mbox{\epsfbox[0 0 260 170]{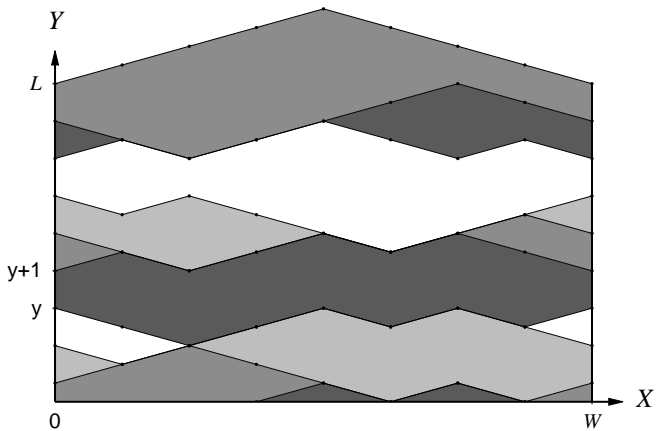}}}
  \caption{
           Schematic representation of optimal RDPs fixed at their two
           endpoints.  An avalanche is defined by the area $S$ spanned
           by the transition from the optimal configuration at $y$ to
           $y+1$, i.e. $S$ is the area interior to the perimeter formed
           by the union of the two optimal RDP configurations at $y$
           and $y+1$ and the two vertical segments $[(0, y);\,(0, y+1)]$
           and $[(W, y);\,(W, y+1)]$. The successive avalanches are
           represented in different grey scales.
          }
  \label{fig2}
\end{figure}

The correspondence between depinning transitions and SOC models \cite{Mapping}
suggests, with the previous results \cite{Mezard}, the following variant of the
problem.  Instead of applying a field (control parameter), we set the depinning
velocity (order parameter) to an infinitesimal value \cite{footn1}.  This is
accomplished by initially fixing the two ends of the polymer at $(x_1=0, y_1=y)$
and $(x_2=W, y_2=y)$.  Since the two ordinates $y_1=y_2=y$ are equal, we could
consider the case where only one endpoint is fixed while keeping the other one
free, therefore making this situation correspond to a polymer on average twice
as long but with both endpoints fixed.   Alternatively, we may consider the
polymer as wrapping itself around a cylinder of circumference $W$.
The polymer is allowed to equilibrate, i.e. take the spatial configuration of
minimum total bond energy.   Let us now shift the vertical position of the
fixed end points from $y$ to $y+1$ (where the lattice mesh is taken as unity). The polymer is again allowed to equilibrate to the spatial configuration of
minimum energy.   We continue in this fashion in an iterative process, which
amounts to controlling the average vertical velocity of the polymer to a value
so small that the time scale to move over a unit mesh is much larger than any
relaxation times.  This guarantees that the polymer always finds the spatial
configuration of minimum bond energy.  Note that the $y$ position of the
polymer end points therefore functions as a clock, since no other relevant time scales are present.

Figure~\ref{fig1} shows a typical set of optimal configurations for a RDP of
length $W=4096$ and for $0 \leq y \leq L = 1200$. The magnifications illustrate
the self-affine structure of the RDPs and the self-similar hierarchical pattern
of the local branching structure.

\section{Theoretical prediction of the avalanche size distribution}
\label{sec:thpr}

In this model, an avalanche at $y$ is simply the transition from the optimal
configuration of a RDP with end points fixed at $y_1=y_2=y$ to the optimal
configuration where the end points are now at $y_1=y_2=y+1$ (as shown in
Fig.~\ref{fig2}).   We define the size of an avalanche by the area $S$ spanned
by the transition from the optimal configuration at $y$ to the one at $y+1$,
i.e. $S$ is the area interior to the perimeter formed by the union of the two
optimal RDP configurations at $y$ and $y+1$ and the two vertical segments
$[(0, y);\, (0, y+1)]$ and $[(W, y);\, (W, y+1)]$ (see Fig.~\ref{fig2}).
What do we know about the distribution of these avalanche sizes?

Clearly, the structure of the ensemble of the optimal RDPs (for all possible
end point $y$ locations) uniquely determines the avalanches.  For a RDP of
length $W$, it is known that the typical transverse excursion $Y$ in $1+1$
dimensions scales as $Y \sim W^{\nu}$ with $\nu = {2\over 3}$ (see
\cite{Halpin-Zhang95} and references therein). We thus expect that there
exists a class of RDP transitions with vertical lengths $Y$ of at least the 
order of this typical transverse excursion. The area $S$ spanned by such a
transition is therefore proportional to $S \simeq WY \sim W^{5\over 3}$ (i.e.
the characteristic avalanche size).  The distribution of $Y$ is known 
to behave asymptotically as  $P(Y) \sim \exp(-({Y \over W^{2\over 3}})^3)$ \cite{Halpin-Zhang95}.   Subtituting for $S \simeq WY$ in $P(Y)$ give us
\be
P(S) \sim \exp(-\biggl({S\over W^{5\over 3}}\biggl)^3)\;,
\label{eq:psgs1}
\ee
for $S$ at least of the order of $W^{5\over 3}$.  This constitutes our first
prediction.  Its validity will be tested numerically in Sec.~\ref{sec:numtst}.

We now derive the distribution of avalanches in the large $W$ limit for $S$
smaller than $W^{5\over 3}$.  First notice that the sequence of optimal paths with fan shaped families of end points strongly resembles the ranking of paths
by Zhang \cite{zhang}.  He found that the difference $Y$ between the endpoints
of these optimal paths scale with path length $X$ as $Y \sim X^{\nu_s}$ with
$\nu_s = {1\over 3}$.  This property is important for the understanding of our
results as it suggests a hierarchical structure.  We thus briefly recall its
derivation.

The Bethe ansatz with the replica trick \cite{Parisi} provides a solution
of the RDP problem in $1+1$ dimension.  This shows that the RDP problem is
equivalent to solving a problem of $n$ bosons in one spatial dimensions
interacting with an attractive delta function potential.   In this framework,
imposing conditions on the endpoints of the RDP implies that the Bethe ansatz
wave function must incorporate the motion of the center of mass of the $n$
bosons:
\be
\Psi \sim 1/\exp(\sum_{\alpha, \beta} |x_{\alpha}-x_{\beta}| +
{1 \over W} \sum_{\alpha = 1}^n x_{\alpha}^2)\;.
\label{eq:bawf1}
\ee
The term ${1\over W} \sum_{\alpha = 1}^n x_{\alpha}^2$ represents the kinetic
and $\sum_{\alpha, \beta} |x_{\alpha}-x_{\beta}|$ the potential energy.  Since
$x_{\alpha} \sim W^{2\over 3}$, the kinetic energy $\sim W^{1\over 3}$.  In 
the Bethe ansatz wave function, the potential energy must be comparable to the
kinetic energy, thus $|x_{\alpha}-x_{\beta}| \sim W^{1\over 3}$,
confirming that $\nu_s = {1 \over 3}$.  This scaling describes the distance
between degenerate ground states with so-called ``weak replica symmetry
breaking'' \cite{Parisi}.  Technically there is a replica symmetry breaking 
but the distance between the degenerate states becomes negligible compared to
their intrinsic fluctuations in the thermodynamic limit.

We generalize the above observations to infer two transverse length scales
$W^{2\over 3}$ and $W^{{2\over 3}\alpha}$, where $0<\alpha\le 1$ (the case
$\alpha=1$ is addressed separetely below), to describe the hierarchical
structure of RDP configurations as exemplified by Fig.~\ref{fig1}.  Our
results turn out to be independent of the choice of $\alpha$.  Intuitively,
a family of width $W^{{2\over 3}\alpha}$ consists of families of width
$W^{{2\over 3}\alpha^2}$ each of which consists of families of smaller width 
and so on (down to the elemental scale of the mesh).  The width, number and
other properties of these embedded sets of families can be obtained from the 
two length scales $W^{2\over 3}$ and $W^{{2\over 3}\alpha}$ using only 
dimension conservation and self-similarity arguments.

\begin{enumerate}
\item The highest order family, that we call of order $1$, corresponds to all
the locally optimal paths that are within a distance of order $W^{2\over 3}$
of a best path.  The vertical width of this family of order $1$ is
$w_1 \propto W^{2\over 3}$.  This family is composed of locally optimal paths
that join after a distance $l_1 \propto W$, obtained by the condition that
$l_1^{2\over 3} \propto W^{2\over 3}$ (this condition will become nontrivial
at lower levels of the hierarchy).  The generic area covered by this family
is $S_1 \propto l_1\, w_1 \sim W^{5\over 3}$.  This is also the typical size
of the largest possible avalanche as defined above and corresponds to a
transition between members of this family of order $1$.

\item Within this family of order $1$, we define $N_2$ families of order $2$,
each of which have a characteristic width
$w_2 \propto l_1^{{2\over 3}\alpha} \sim W^{{2\over 3}\alpha}$.  It is at
this point that we have used the second length scale introduced by the
quasi-degenerate ground states.  From the conservation of (vertical) width,
we have by construction,
\be
N_2 w_2 = w_1\;,
\label{eq:cowi1}
\ee
leading to $N_2 \propto W^{{2\over 3}(1-\alpha)}$.   A family of order $2$
is by itself composed of locally optimal paths that join after a distance
$l_2 \propto W^{\alpha}$, obtained by the self-consistent condition that
\be
w_2 \propto l_2^{2\over 3}\;.
\label{eq:wops1}
\ee
As a consequence, the generic area, i.e. the largest possible avalanche,
covered by this family of order $2$ (intra member transitions) is
$S_2 \propto l_2 ~w_2 \sim W^{\alpha} \times W^{{2\over 3}\alpha} =
W^{{5\over 3}\alpha}$.

\item[{\it n}.] We infer that the relevant quantities of the $n$-th order family
only depends on the associated ones in the family of order $n-1$.  This
leads us to a recursive scheme for the calculation of the above introduced
entities.  In what follows we will formally define the simplest version of
the iterative system of equations and state its solutions.

\end{enumerate}

Within each of the families of order $n$, we define $N_{n+1}$ families of
order $n+1$, each of which have a characteristic (vertical) width $w_{n+1}$.
From the conservation of width, we have by construction,
\be
N_{n+1} w_{n+1} = w_n\;.
\label{eq:cowin}
\ee
The characteristic width $w_{n+1}$ relates the generic distance $l_n$ after
which locally optimal paths (within a family of order $n$) typically join.
It obeys
\be
w_{n+1} = a l_n^{{2\over 3}\alpha}\;.
\label{eq:wnpln}
\ee
But the self-consistency condition relates $w_{n+1}$ to $l_{n+1}$ 
\be
w_{n+1} = B l_{n+1}^{2\over 3}\;.
\label{eq:wopsn}
\ee
We are thus led to the direct recursion
\be
l_{n+1} = A l_n^{\alpha}\;.
\label{eq:lnpln}
\ee
The typical area covered by an avalanche among the families of order $n+1$, is
\be
S_{n+1} = C l_{n+1} w_{n+1}\;.
\label{eq:snplw}
\ee
Here $A$, $a$, $B$, and $C$ are (real valued) constants.  Since $l_1 \propto W$,
we also have an initial condition for the recursion.  This is generalized as 
\be
l_1 = f(W)\;.
\label{eq:init1}
\ee
Finally, we have for the total number of families ${\cal N}_{n+1}$ up to
and including order $n+1$,
\ba
{\cal N}_{n+1} &=& N_{n+1} {\cal N}_n\;, \label{eq:nnpn1}\\
{\cal N}_1 &=& 1\;.
\label{eq:nnpn2}
\ea
We find from Eqs.~(\ref{eq:lnpln}) and (\ref{eq:init1})
\be
l_n = A^{1\over 1-\alpha}2 \,
  \biggl({f(W)\over A^{\alpha\over 1-\alpha}}\biggl)^{\alpha^{n-1}}
  \quad\mbox{for}\quad  n \ge 1\;,
\label{eq:lnrw1}
\ee
and  from Eq.~(\ref{eq:wopsn}) that
\be
w_n = B A^{2\over 3(1-\alpha)} \,
\biggl({f(W)\over A^{\alpha\over 1-\alpha}}\biggl)^{{2\over 3}\alpha^{n-1}}\;.
\label{eq:worw1}
\ee
Together with Eq.~(\ref{eq:cowin}) we then get,
\be
N_{n+1} =
\biggl({f(W)\over A^{\alpha\over 1-\alpha}}\biggl)^{{2\over
3}(1-\alpha)\alpha^{n-1}}\;,
\label{eq:cownpw}
\ee
and from Eq.~(\ref{eq:snplw}),
\be
S_n = C ~B ~^{5\over 3(1-\alpha)} \,
\biggl({f(W)\over A^{\alpha\over 1-\alpha}}\biggl)^{{5\over 3}\alpha^{n-1}}\;.
\label{eq:snwnn}
\ee
The latter expression is conveniently turned into
\be
\alpha^{n-1} =
{3\over 5} \ln \biggl({S_n\over {C B A^{5\over 3(1-\alpha)}}}\biggl)/
   \ln \biggl({f(W)\over A^{\alpha\over 1-\alpha}}\biggl)\;.
\label{eq:twonm}
\ee
For the cumulative number of families to order $n$, i.e. Eqs.~(\ref{eq:nnpn1})
and (\ref{eq:nnpn2}), we get with Eq.~(\ref{eq:cownpw})
\be
{\cal N}_n =
  \biggl({f(W)\over A^{\alpha\over 1-\alpha}}\biggl)^{{2\over
3}{(1-\alpha^{n-1})}}\;,
\label{eq:nnwn1}
\ee
which with Eq.~(\ref{eq:twonm}) results in a direct $S_n$ and $W$ dependence,
\be
{\cal N}_n = (C B)^{2\over 5} f(W)^{2\over 3}/S_n^{2\over 5}\;.
\label{eq:nnsnw}
\ee

This reasoning, based on the hierarchical model, gives us the number of
avalanches of specific sizes.  To get the probability density distribution, 
we have to divide this number by the interval width from $S_n$ to $S_{n+1}$
which is simply proportional to $S_n$ up to a correction of order
$S_n^{\alpha-1}$ as seen from Eq. (\ref{eq:snwnn}).  Gathering all the 
pieces and assuming that $f(W) \propto W$ leads us to the following prediction
for the distribution of avalanche sizes
\be
P(S)\,dS \propto {W^{2\over 3}\over S^{1 + \mu}}\,dS\;,
\label{eq:psspm}
\ee
with an exponent
\be
\mu = {2\over 5}\;.
\label{eq:muex1}
\ee
The power law in Eq. (\ref{eq:psspm}) describes the distribution of avalanche
sizes $1 \leq S \leq S_{5\over 3}$ (we define
$S_{5\over 3} \propto W^{5\over 3}$) in the limit $W \to \infty$.  This upper
scale $S_{5\over 3}$ corresponds to the maximum typical sizes of the avalanches
of order $1$ in the hierarchy.  Notice that the prediction of
Eq.~(\ref{eq:muex1}) is independent of the value $0 < \alpha < 1$ and is thus robust with respect to the detailed structure of the hierarchy.

A similar hierarchical structure can also be constructed for $\alpha=1$.  In
this case, it is postulated that $w_{n+1}=w_{n}/\lambda$, where $\lambda>1$
is the {\it constant} reduction factor from one level of the hierarchy to the
next.  While keeping Eq.~(\ref{eq:cowin}), this leads to
$l_{n+1}=l_{n}/\lambda^{3\over 2}$ and thus to
$S_{n+1}=S_{n}/\lambda^{5\over 2}$ using Eq.~(\ref{eq:snplw}).   The total
number of families of order $n$ is now simply proportional to $\lambda^{n}$.
Solving as a function of $S_{n}$, we retrieve exactly expression
(\ref{eq:psspm}) for the distribution of avalanche sizes.

This derivation is simpler because the hierarchical structure is exactly
self-similar, with the same scaling ratio $\lambda$ throughout.  This is
in contrast to the case $\alpha<1$ for which
$\lambda \propto w^{{2\over 3}\alpha^{n-1}(1-\alpha)}$ decreases with 
increasing family order.  This derivation for $\alpha=1$ and $\lambda>1$
clarifies the origin of the exponent $\mu={2/5}$ stemming simply from
${1/\mu}=1+{3/2}$, i.e. from the fundamental self-affine structure
of the RDP with transverse excursion exponent ${2/3}$.

We thus stress that the prediction of Eqs.~(\ref{eq:psspm}) and 
(\ref{eq:muex1}) is very general and independent of the specific hierarchical
structure of the sub-dominant quasi-degenerate ground states.  Note that the
power-law distribution given by Eq.~(\ref{eq:psspm}) for the spanned surfaces 
is associated with two other power-laws, namely that for the distribution of
typical transverse deviations $w$ and that for the distribution of typical
longitudinal deviations $l$.  This stems from $S \sim w\,l$ and 
$w \sim l^{2/3}$, leading to
\be
P(w)\,dw \sim {dw\over w^{1 + 1}}\quad\mbox{and}\quad
P(l)\,dl \sim {dl\over l^{1 + 2/3}}\;.
\label{eq:plslc}
\ee
As mentioned in Sec.~\ref{sec:intro}, our finding seem to be in disagreement
with the predicted value (equal to zero) for the avalanche size distribution
in $1+1$ dimensions \cite{middle,Nofish,Hwa}.  However, the avalanche regime
of interest to us is different from that previously investigated.  Our regim
consists of small and intermediate avalanches of sizes up to  $S_{5\over 3}$,
whereas the regime which contains avalanches larger than $S_{5\over 3}$ yields 
a size distribution with a vanishing exponent.  The present work proposes a
sub-dominant power-law distribution of avalanches which originates in a
hierarchical ordering of the almost equivalent degenerate states.  Previous 
work has addressed the tail end of the avalanche distribution and has therfore
not been attentive to the presence of these states.

\section{Numerical tests}
\label{sec:numtst}

\begin{figure}
  \centerline{\mbox{\epsfbox[0 0 244 244]{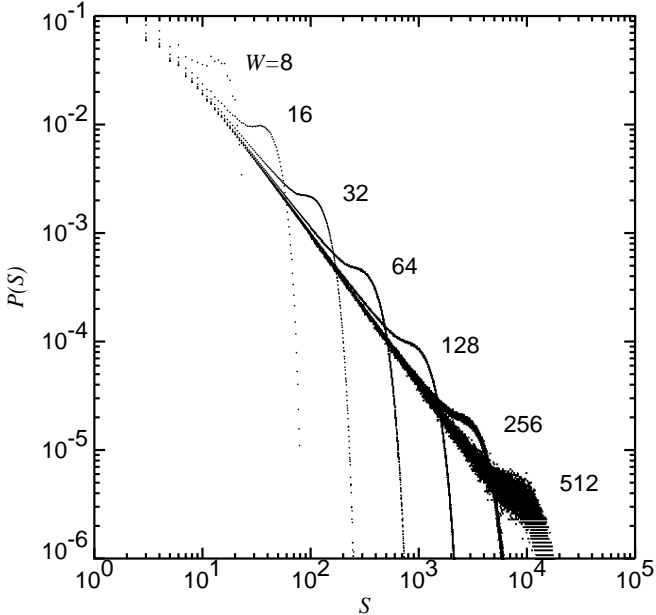}}}
  \caption{
           Distribution $P(S)$ of RDP avalanche sizes obtained numerically
           for system widths from $W = 8$ to $512$ in a log-log plot.
           Here the system lengths $L$ are $2\times 10^7$ (for $W=8$),
           $3\times 10^6$ ($16$), $2\times 10^7$ ($32$), $10^8$ ($64$),
           $2\times 10^8$ ($128$), $5\times 10^7$ ($256$), and
           $9\times 10^6$ ($512$).
          }
  \label{fig3}
\end{figure}

The distribution of avalanche sizes $S$ has been determined numerically by
using a now standard transfer matrix method \cite{tranfer} relying on the
chain property applying to the energy $e(x_1,y_1;x_2,y_2)$ of a RDP going
from $(x_1,y_1)$ to $(x_2,y_2)$:
\be
e(x_1,y_1;x_2,y_2)=\min_{y'}[e(x_1,y_1;x',y') + e(x',y';x_2,y_2)]\;.
\label{eq:mmetm}
\ee
Figure~\ref{fig3} shows the distribution of avalanche sizes obtained
numerically for system widths from $W = 8$ to $512$.  For each width,
we have calculated the RDP configurations and the corresponding avalanche
areas for system lengths $3\times 10^6 \le L \le 2\times 10^8$.  These
very long systems provide reliable statistical estimates.
In Fig.~\ref{fig3}, the existence of a power-law for the distribution
$P(S)$ is quite apparent.  The size interval over which the power-law
holds increases as $S_{5\over 3} \sim W^{5/3}$.
Another feature of Fig.~\ref{fig3} is the clear evidence of a characteristic
avalanche size, corresponding to the bump of the distribution in the region 
of large avalanche sizes.  The location of these bumps also scales as
$S_{5\over 3} \sim W^{5/3}$.
Finite size effects turn out to be very important in this problem and a 
careful finite size scaling analysis is appropriate.  We approach this as
follows.  
\begin{figure}
  \centerline{\mbox{\epsfbox[0 0 244 282]{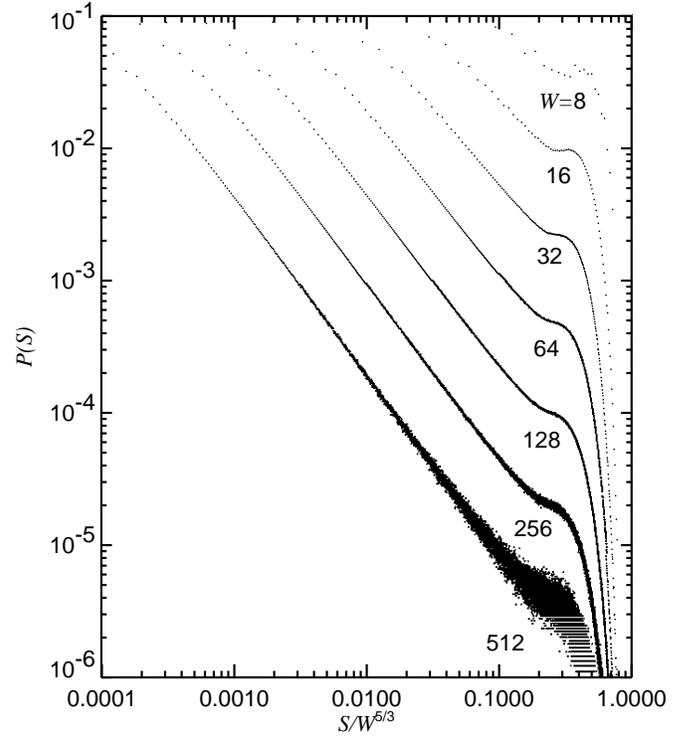}}}
  \caption{
           $P(S)$ as a function of the rescaled variable $S/W^{5/3}$
           for $W = 8$ to $512$ in a log-log plot.
          }
  \label{fig4}
\end{figure}
\begin{figure}
  \centerline{\mbox{\epsfbox[0 0 244 216]{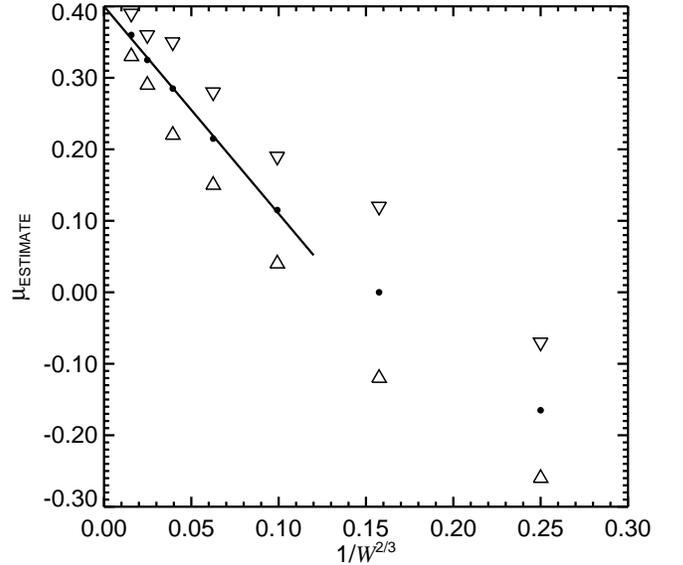}}}
  \caption{
           The estimated $\mu$'s dependence on $W^{-2/3}$.  These $\mu$
           values are the result of a linear fit of $1/S^{1+\mu}$ to the
           linear portions in Fig.~\ref{fig4}.  The high, low, and midpoint
           estimates are indicated by $\bigtriangledown$, $\bigtriangleup$,
           and $\bullet$ respectively.  The straight line is the least
           squares fit to the midpoint values with the five largest system
           widths ($W=32$, $64$, $128$, $256$, $512$).  This line has
           been extended to the $W\to \infty$ limit.
          }
  \label{fig5}
\end{figure}
For each system size, we determine the exponent $\mu(W)$ which 
best fits the numerical distribution.  To demonstrate the quality of the 
fit, we replot Fig.~\ref{fig3} by showing in Fig.~\ref{fig4} the function 
$P(S)$ as a function of the rescaled variable $S/W^{5\over 3}$.  For each 
size $W$, a different exponent $\mu(W)$ is found.  The dependence of 
$\mu(W)$ as a function of $W^{-{2\over 3}}$ is shown in Fig.~\ref{fig5}.  
We find a very good fit (``least squares'') with the finite size equation
\be
\mu(W) = \mu_{\infty} - {c\over W^{2\over 3}}\;,
\label{eq:mufit}
\ee
where $c=2.90$ is a constant and $\mu_{\infty} =0.40$.  This is in excellent
agreement with the prediction $2/5$.  An apparent power-law dependence of
the exponent $\mu(W)$ on $W$ as in Eq.~(\ref{eq:mufit}) could result from
fluctuations in the value of $\alpha$ within each level and accross the
different levels of the hierarchy.
\begin{figure}
  \centerline{\mbox{\epsfbox[0 0 244 180]{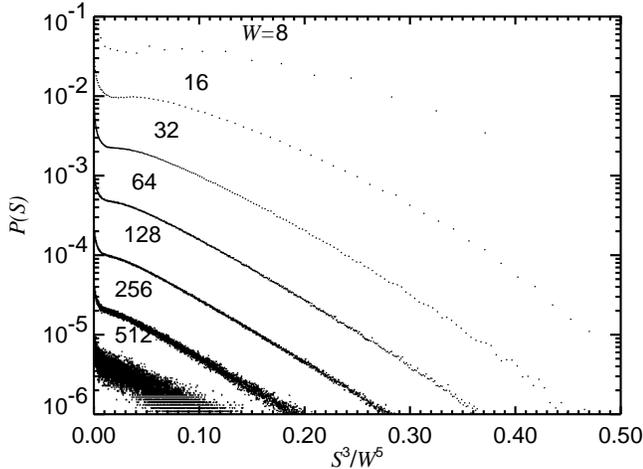}}}

  \caption{
           $P(s)$ for the different system sizes as a function of the
           rescaled variable  $S^{3}/W^{5}$ in a semilog plot.
          }
  \label{fig6}
\end{figure}
Finally, in Fig.~\ref{fig6} we represent
$P(s)$ for the different system sizes as a function of the rescaled
variable  $S^{3}/W^{5}$.  This choice of variables is intended to test the
prediction of Eq.~(\ref{eq:psgs1}).  We observe a rather convincing tendency 
for the plot to converge to a straight line for large system sizes.

\section{Conclusion}
\label{sec:conc}

We have proposed a novel quasi-statically driven model that exhibits responses
similar to those of SOC models.  This model of a succession of optimal RDP
configurations exhibits a power-law distribution of the area swept by a polymer
between two successive optimal configurations (defined as an avalanche).

Based on the existence of two fundamental scales $W^{2\over 3}$ and
$W^{{2\over 3}\alpha}$ ($0<\alpha\le 1$) for the transverse fluctuations of a
RDP of length $W$, we have constructed a hierarchical representation of the
set of quasi-degenerate optimal configurations.  This hierarchy allow us to
calculate explicitely the exponent of the avalanche distribution.

Our numerical analysis confirm the existence of two distinct populations of
avalanches.  One of these populations consists of ``small'' avalanches that
are distributed according to a power law with an upper cutoff controlled by
the typical transverse length scale $W^{2\over 3}$.  The other population
comprises the ``large'' avalanches beyond this typical transverse excursion
$W^{2\over 3}$.

\acknowledgements

We are grateful to M.~M\'ezard and Y.-C.~Zhang for stimulating discussions
and I.~Dornic for help in the initial stage of this work.

%
%

%
%
%
\end{multicols}


\begin{references}
\bibitem{footn1}
Usually, the depinning velocity would be controlled by friction and
relaxation effects not discussed here; it turns out that these
ingredients which are essential for describing the dynamical regime
are not necessary in the SOC regime where one controls the velocity to
an infinitesimal value, since this ensures that the problem consists
of a succession of (at least locally) equilibrium problems.
\end{references}
\end{document}